\documentclass{article}
\usepackage{spconf,amsmath,graphicx,hyperref}
\usepackage{makecell}
\usepackage{multirow}
\usepackage{booktabs}
\usepackage{amsfonts}
\usepackage{cite}
\usepackage{bm}
\usepackage{svg}
\usepackage{gensymb}
\usepackage{circledsteps}
\usepackage[subtle,title=normal,sections=normal,margins=normal,mathdisplays=normal,mathspacing=normal, bibliography=normal]{savetrees}
\usepackage{blindtext}
\usepackage{xcolor}

\title{Geometry-Conditioned Target Speaker Extraction \\ based on spatial features incorporating acoustic propagation principles}
\name{Jiatong Li, Simon Doclo\thanks{This work was funded by the Deutsche Forschungsgemeinschaft (DFG, German Research Foundation) under Germany's Excellence Strategy - EXC 2177/2 - Project ID 390895286, Project ID 352015383 - SFB 1330 B2, and Project ID 568930428. Simulations were conducted on the HPC cluster ROSA funded by the DFG under INST 184/225-1 FUGG.}}
\address{Dept. of Medical Physics and Acoustics and Cluster of Excellence Hearing4all,\\ Carl von Ossietzky Universität Oldenburg, Oldenburg, Germany\\
\{\href{mailto:jiatong.li@uni-oldenburg.de}{jiatong.li}, \href{mailto:simon.doclo@uni-oldenburg.de}{simon.doclo}\}@uni-oldenburg.de}
\begin{document}
\ninept
\maketitle
\begin{abstract}
Recently, spatially selective non-linear filter (SSF) was proposed for target speaker extraction, utilizing the direction-of-arrival (DOA) of the target speaker as a spatial cue. To overcome its dependency on a fixed microphone array geometry, geometry-conditioned SSF (GCSSF) was proposed, which maps the target DOA and the microphone positions into abstract high-dimensional embeddings using positional encoding. As an alternative to these high-dimensional embeddings, in this paper we propose utilizing two spatial features incorporating acoustic propagation principles for geometry conditioning: 1) the steering vector of the target speaker, which captures inter-microphone phase differences, and 2) a spatial mask based on GCC-PHAT, which quantifies presence of the target speaker per time-frequency bin. These features are used as input to a geometry-conditioning encoder and integrated via a FiLM layer into the GCSSF system. Experimental results on circular and random arrays demonstrate that conditioning with spatial features improves target extraction performance compared to abstract high-dimensional embeddings, maintaining a clear performance advantage across a wide range of target DOA errors.
\end{abstract}
\section{Introduction}
\label{sec:intro}
Extracting a target speaker from a mixture of speakers and background noise remains a fundamental challenge in acoustic signal processing \cite{overview}. To discriminate the target speaker from interfering sources, various cues have been proposed, such as enrollment utterances \cite{10889409,11226272}, visual information \cite{9380418}, and spatial features \cite{beamformer_guided, tesch2024ssf, 11230948, shetu2025ganbasedmultimicrophonespatialtarget}. In this paper, we focus on the spatially selective non-linear filter (SSF) proposed in \cite{tesch2024ssf}. The SSF utilizes the target direction-of-arrival (DOA) as a spatial cue and uses Long Short-Term Memory (LSTM) layers to estimate a complex-valued mask that is applied to the reference microphone signal to extract the target speaker.

Despite its effectiveness, a limitation of the standard SSF is its inherent dependency on the fixed microphone array geometry considered during training. Recently, geometry-agnostic systems have been proposed for noise reduction \cite{jukic2023TAC, Tammen2024TAC, eigenbeam2026, liu2026arrayinvariantspeechenhancementgeometryaware}, blind speaker separation \cite{flexio2026}, and target speaker extraction \cite{AmbiDrop2026, li2026flexiblemultichanneltargetspeaker}. While the system in \cite{AmbiDrop2026} is restricted to a fixed look direction, the geometry-conditioned SSF (GCSSF) system in  \cite{li2026flexiblemultichanneltargetspeaker} allows for arbitrary DOA-steered extraction. The GCSSF system utilizes microphone positional encoding (MPE) \cite{pe} to map the target DOA and microphone positions into abstract high-dimensional sinusoidal embeddings.
\begin{figure}[t]
  \centering
    \includegraphics[width= 0.97\linewidth]{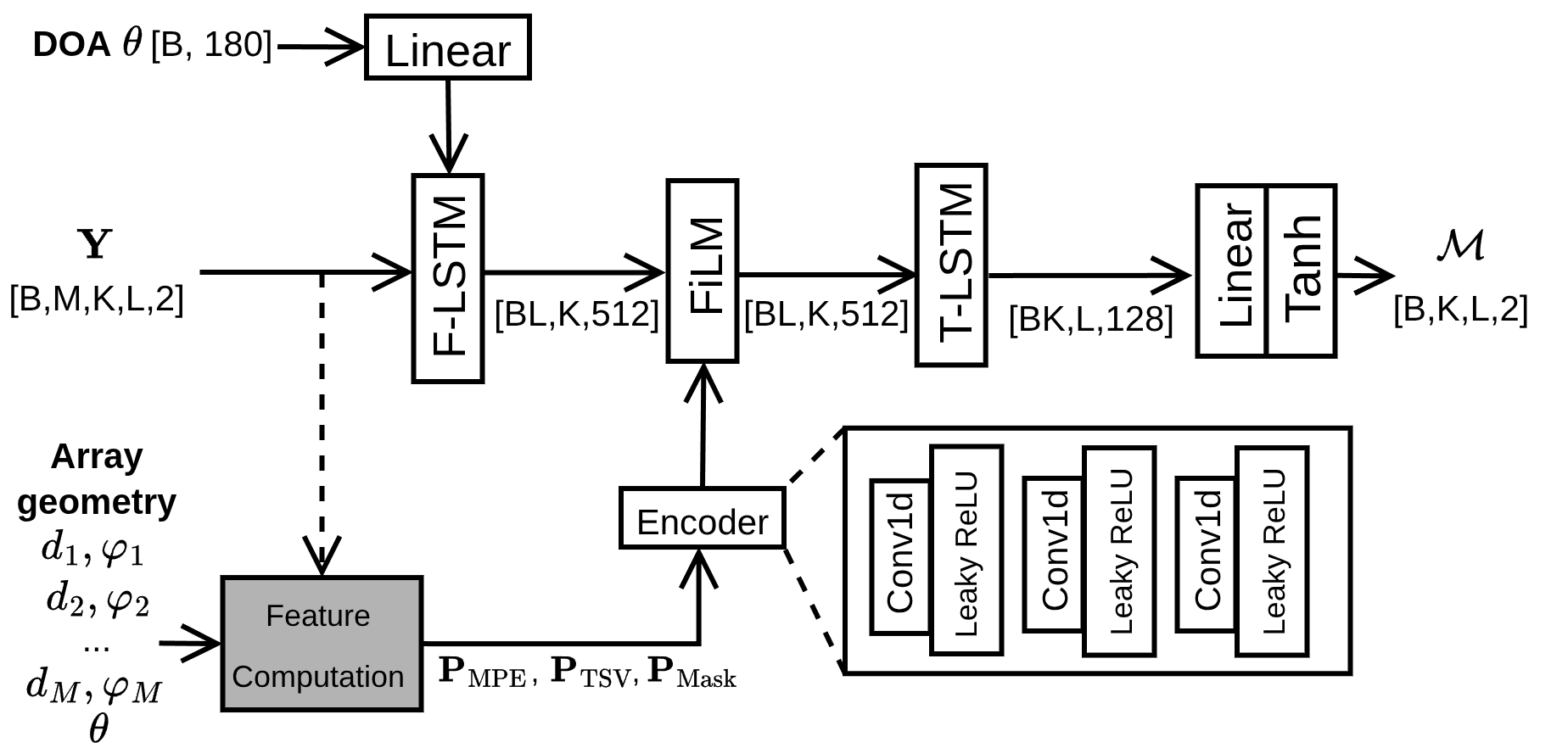}
  \label{fig:system}
\vspace{-2ex}
 \caption{Geometry-conditioned spatially selective filter (GCSSF) system using different features for array geometry conditioning. The gray block highlights the computation of these features ($\mathbf{P}_{\text{MPE}}$, $\mathbf{P}_{\text{TSV}}$, and $\mathbf{P}_{\text{Mask}}$). The features are encoded and then integrated via a FiLM layer.}
\label{fig:diagram}
\end{figure}

As an alternative to these abstract embeddings, in this paper we propose utilizing two spatial features for geometry conditioning, which incorporate acoustic propagation principles into the conditioning mechanism (see Fig.~\ref{fig:diagram}). The spatial features are: 1) the target steering vector (TSV), calculated from the time differences of arrival, which captures the phase shifts of the target signal across the microphones; and 2) a spatial mask based on the generalized cross-correlation with phase transform (GCC-PHAT) \cite{7074215}, which quantifies target presence per time-frequency bin. Similar to the MPE features in \cite{li2026flexiblemultichanneltargetspeaker}, these features are used as input to the geometry-conditioning branch of the GCSSF, which includes an encoder and a Feature-wise Linear Modulation (FiLM) layer \cite{film}. In the experimental evaluation, the proposed GCSSF systems using the proposed spatial features as well as the baseline SSF and the GCSSF-MPE system are trained on random arrays and evaluated on both circular and random arrays. Experimental results demonstrate that conditioning the network with explicit spatial features consistently outperforms the baselines across all evaluated metrics, with the TSV feature yielding the best performance. Furthermore, a sensitivity analysis demonstrates that the proposed systems maintain better target speaker extraction performance than the baseline systems for target DOA errors up to $20^\circ$.

\section{Geometry-conditioned Spatially Selective Non-linear Filter}
\label{sec:ssf}
In this section, we review the geometry-conditioned spatially selective non-linear filter (GCSSF) for target speaker extraction \cite{li2026flexiblemultichanneltargetspeaker}. We consider an acoustic scenario with $J$ speakers and background noise in a reverberant room, where one of the speakers is considered as the target speaker. The sound sources are recorded by an arbitrary planar array with $M$ omni-directional microphones, where the polar coordinates of the $m$-th microphone relative to the array centroid are denoted by  $(d_m, \varphi_m)$.

In the short-time Fourier transform (STFT) domain, the observed noisy speech signal at the $m$-th microphone for frequency bin $k \in [1, K]$ and time frame $l \in [1, L]$ is denoted by $Y_m(k,l)$, where $K$ and $L$ denote the number of frequency bins and time frames, respectively. By stacking the signals from all microphones, the noisy speech vector $\mathbf{Y}(k,l) = [Y_1(k,l), Y_2(k,l), ... , Y_M(k,l)]^\text{T} \in \mathbb{C}^M$ is given by
\begin{equation}
\mathbf{Y}(k,l) = \mathbf{X}(k,l) + \mathbf{V}(k,l),
\end{equation}
where $\mathbf{X}(k,l) \in \mathbb{C}^M$ denotes the target speech vector and $\mathbf{V}(k,l) \in \mathbb{C}^M$ represents the combination of interfering speakers and background noise. Both vectors are defined similarly to $\mathbf{Y}(k,l)$.

The goal is to recover the reverberant target speech component $X_1(k,l)$ from the noisy speech vector $\mathbf{Y}(k,l)$, with the first microphone chosen as the reference microphone without loss of generality. To achieve this, the GCSSF estimates a complex mask $\mathcal{M}(k,l)$, which is applied to the noisy reference microphone signal $Y_1(k,l)$, i.e.,
\begin{equation}
\hat{X}_1(k,l) = \mathcal{M}(k,l) Y_1(k,l).
\end{equation}

Fig.~\ref{fig:diagram} depicts the block diagram of the GCSSF system, which consists of an extraction branch and a geometry-conditioning branch. The extraction branch has the same structure as the baseline SSF system \cite{tesch2024ssf}, i.e., a frequency-domain LSTM (F-LSTM), a time-domain LSTM (T-LSTM), and a linear layer to jointly process spatial, spectral, and temporal information. In addition, the target DOA $\theta$ is used as a conditioning input to the F-LSTM layer.
To enable generalization to different microphone array geometries, the geometry-conditioning branch processes features based on the array geometry by an encoder, whose output is utilized by a FiLM layer after F-LSTM to modulate the intermediate representations within the extraction branch. In \cite{li2026flexiblemultichanneltargetspeaker}, a microphone positional encoding (MPE) scheme directly maps the microphone coordinates and the target DOA into high-dimensional sinusoidal embeddings without requiring learnable parameters. The positional encodings for the $m$-th microphone and the target DOA are given by
\begin{equation}\mathbf{p}_m = d_m \begin{bmatrix} \cos(2\pi\sigma \mathbf{v} + \varphi_m) \\ \sin(2\pi\sigma \mathbf{v} + \varphi_m) \end{bmatrix} \in \mathbb{R}^Q,
\end{equation}
\begin{equation}
\mathbf{p}_\text{DOA} = \begin{bmatrix} \cos(2\pi\sigma \mathbf{v} + \theta) \\  \sin(2\pi\sigma \mathbf{v} + \theta) \end{bmatrix} \in \mathbb{R}^Q,
\end{equation}
where $\mathbf{v} = \frac{2}{Q}[0, 1, ..., \frac{Q}{2}-1]^\text{T}$, and $\sigma$ and $Q$ are constant hyperparameters.
The MPE feature $\mathbf{P}_{\text{MPE}}$ is formed by concatenating the positional encodings for all microphones with the positional encoding for the target DOA, i.e.,
\begin{equation}
\boxed{\mathbf{P}_{\text{MPE}} = [\mathbf{p}_1, \mathbf{p}_2, \dots, \mathbf{p}_M, \mathbf{p}_{\text{DOA}}] \in \mathbb{R}^{Q \times (M+1)}}
\label{eq:mpe}
\end{equation}
For a given batch size $B$, the MPE feature is reshaped to $[B, M+1, Q]$ before being passed to the geometry-conditioning branch.

While this feature provides the network with spatial coordinates, it does not explicitly utilize spatial features derived from acoustic propagation principles. Consequently, the network may need to implicitly learn the relationship between inter-microphone phase differences, the target DOA and the array geometry, potentially limiting extraction performance.
\section{Explicit Spatial Features for \\ Geometry Conditioning}
\label{sec: gc-ssf}
To address the potential limitations of the positional encodings, in this paper we propose utilizing two explicit spatial features incorporating acoustic propagation principles for geometry conditioning. The first feature is the target steering vector (section~\ref{sec:tsv}), whereas the second feature is a spatial mask based on GCC-PHAT (section~\ref{sec:mask}).
\subsection{Target Steering Vector}
\label{sec:tsv}
To capture the physical relationship between the target speaker and the microphone array, the target steering vector is computed from the array geometry and the target DOA.
Assuming a far-field propagation model, the time difference of arrival (TDOA) between the $m$-th microphone and the centroid of the microphone array for a target speaker at DOA $\theta$ is given by $\tau_m = \frac{d_m\cos(\varphi_m - \theta)}{c}$, where $c$ is the speed of sound. Relative to the reference microphone, the target steering vector at frequency $f_k$ for the $k$-th frequency bin is given by:
\begin{equation}
    \boxed{\mathbf{d}(k)=[1, e^{\jmath 2\pi f_k (\tau_2 - \tau_1)},..., e^{\jmath2\pi f_k (\tau_M - \tau_1)}]^\text{T} \in \mathbb{C}^M}
    \label{eq:tsv}
\end{equation}
where $\jmath$ denotes the imaginary unit. This vector describes the frequency-dependent phase differences across the array for the direct path of the target speaker. As is well known from array signal processing \cite{665}, this steering vector can be used in a delay-and-sum beamformer that spatially aligns and extracts the target speaker via the linear operation $\mathbf{d}^H\mathbf{Y}$. This demonstrates that the complex conjugate of the steering vector, $\mathbf{d}^*(k)$, defines the phase compensation required to spatially align the microphone signals to the target DOA. Our first geometry-conditioning feature $\mathbf{P}_{\text{TSV}}$ is constructed by computing and stacking $\mathbf{d}^*(k)$ across all $K$ frequency bins. This yields a tensor of shape $[B, M, K]$, explicitly guiding the neural network to focus on the target DOA.
\subsection{GCC-PHAT Spatial Mask}
\label{sec:mask}
The target steering vector provides a purely geometric feature because it only relies on the microphone coordinates and the target DOA. It is calculated independently of the observed multichannel microphone signals, meaning it cannot capture target speaker activity. To explore whether the network benefits from a signal-dependent spatial feature that indicates the presence of the target speaker in each time-frequency bin, we propose to utilize the GCC-PHAT spatial mask \cite{7074215} as an alternative conditioning feature. This feature is conceptually related to classical spectral post-filtering \cite{Gannot2008}.

To construct the spatial mask, the microphone signals are first spatially aligned toward the target DOA by applying the complex conjugate of the target steering vector in (\ref{eq:tsv}). The TDOA-compensated signal, denoted as $\mathbf{S}(k,l) = [S_1(k,l), S_2(k,l), ... , S_M(k,l)]^\text{T} \in \mathbb{C}^M$, is computed as
\begin{equation}
\mathbf{S}(k,l) = \mathbf{d}^*(k) \odot \mathbf{Y}(k,l), 
\end{equation}
where $\odot$ denotes element-wise multiplication.
Subsequently, for any given pair of TDOA-compensated microphone channels $i$ and $j$, the instantaneous GCC-PHAT $\Psi_{ij}(k,l)$ is computed as
\begin{equation}
    \Psi_{ij}(k,l) = \frac{S_{i}(k,l) S_{j}^*(k,l)}{|S_{i}(k,l)||S_{j}(k,l)|}. 
\end{equation}
To aggregate these phase correlations across the entire array, the GCC-PHAT spatial mask $I(k,l)$ is obtained by averaging the real part of $\Psi_{ij}(k,l)$ over all $\frac{M(M-1)}{2}$ unique microphone pairs, i.e.,
\begin{equation}
    \boxed{I(k,l) = \frac{2}{M(M-1)} \text{Re} \left( \sum_{i=1}^M \sum_{j=i+1}^M \Psi_{ij}(k,l) \right)}
\end{equation}
Theoretically, $I(k,l)$ serves as an indicator of the target presence within a specific time-frequency bin. When the target speaker is the dominant active source, the aligned signals $S_i(k,l)$ and $S_j(k,l)$ are in-phase, causing $I(k,l)$ to approach $1$. Conversely, when a time-frequency bin is dominated by interfering speakers from a different DOA than the target speaker, reverberation or noise, the phase alignment degrades, and $I(k,l)$ decreases. The second geometry-conditioning feature, $\mathbf{P}_{\text{Mask}}$, aggregates $I(k,l)$ across the $K$ frequency bins for each individual time step. To ensure online processing, the network processes this feature frame-by-frame, which corresponds to an input tensor of shape $[BL, 1, K]$ where all $L$ frames are treated independently.

\section{Experiments}
This section first presents the experimental setup, including the training and evaluation datasets, the network structure, and the training procedure.
Then, the experimental results are presented and discussed, evaluating the target speaker extraction performance of the GCSSF systems utilizing the explicit spatial features for geometry conditioning. The performance is compared against two baseline systems: the original SSF system without geometry conditioning \cite{tesch2024ssf} and the GCSSF system using the MPE feature in (\ref{eq:mpe}) for geometry conditioning (GCSSF-MPE) \cite{li2026flexiblemultichanneltargetspeaker}.
\subsection{Datasets}
\begin{figure}[t]
    \centering
    \begin{minipage}[b]{0.55\linewidth}
        \hspace{-7mm}
        \includegraphics[width=4.5cm]{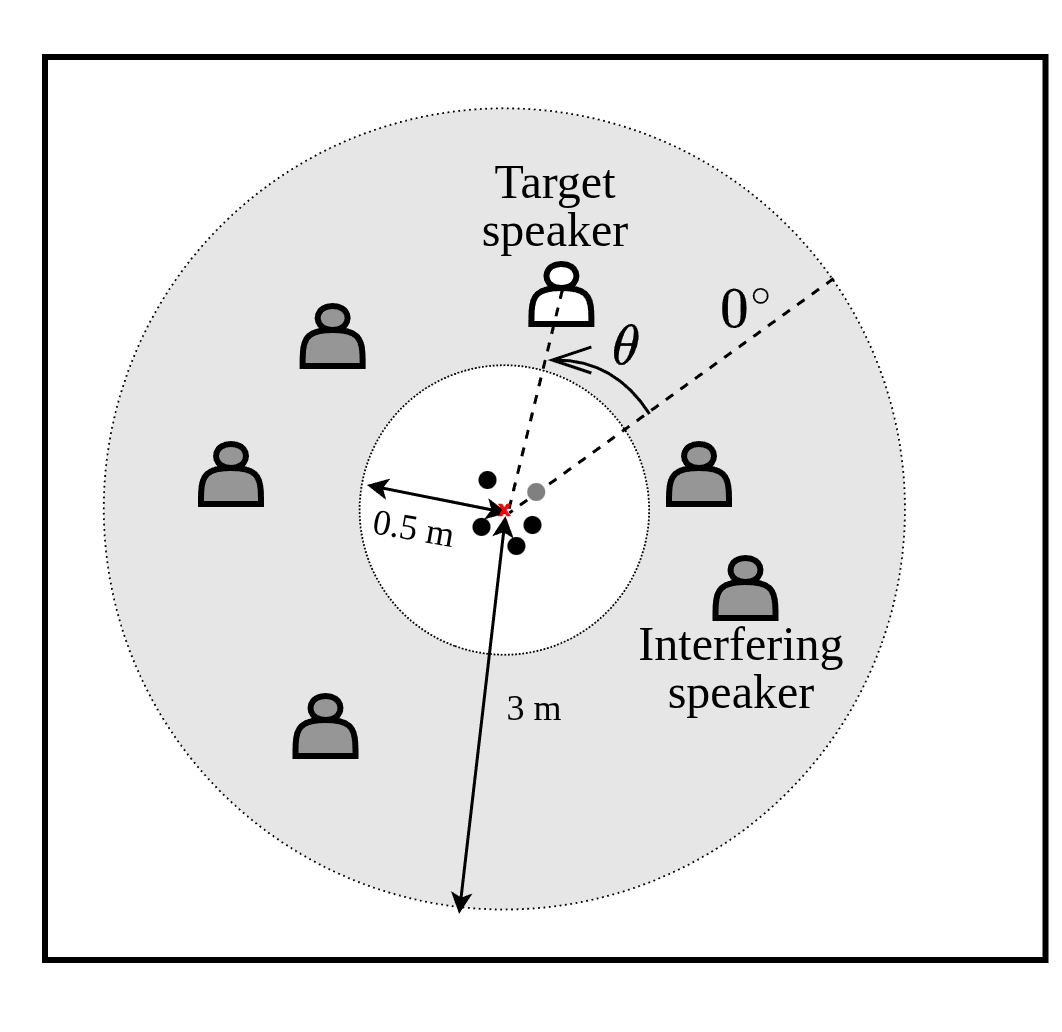}
    \end{minipage}
    \begin{minipage}[b]{0.3\linewidth}
        \hspace{-6mm}
        \small 
        \begin{tabular}{ll}
            \toprule
            \multicolumn{2}{c}{\textbf{Room characteristics}} \\ 
            \midrule
            Width  & 2.5--5 m \\
            Length & 3--9 m   \\
            Height & 2.2--3.5 m \\
            $T_{60}$    & 0.2--0.5 s \\
            \bottomrule
        \end{tabular}
        \vspace{1cm}
    \end{minipage}
    \vspace{-3mm}
    \caption{Illustration of the simulation setup and the room characteristics. The speakers are located in the gray region.}
    \label{fig:setups}
\end{figure}

\begin{figure}[t]
\centering
\begin{minipage}[b]{0.30\linewidth}
  \centering
  \centerline{\includegraphics[width=2.5cm]{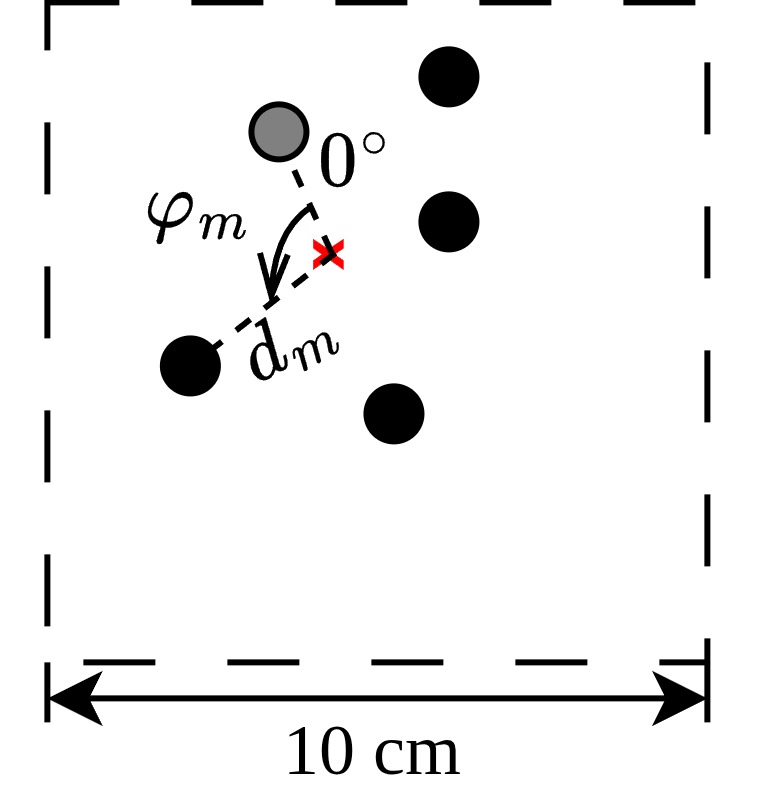}}
 \vspace{-0.1cm}
  \centerline{(a) Random}
  \label{figa}
\end{minipage}
\begin{minipage}[b]{0.30\linewidth}
  \centering
  \centerline{\includegraphics[width=2.4cm]{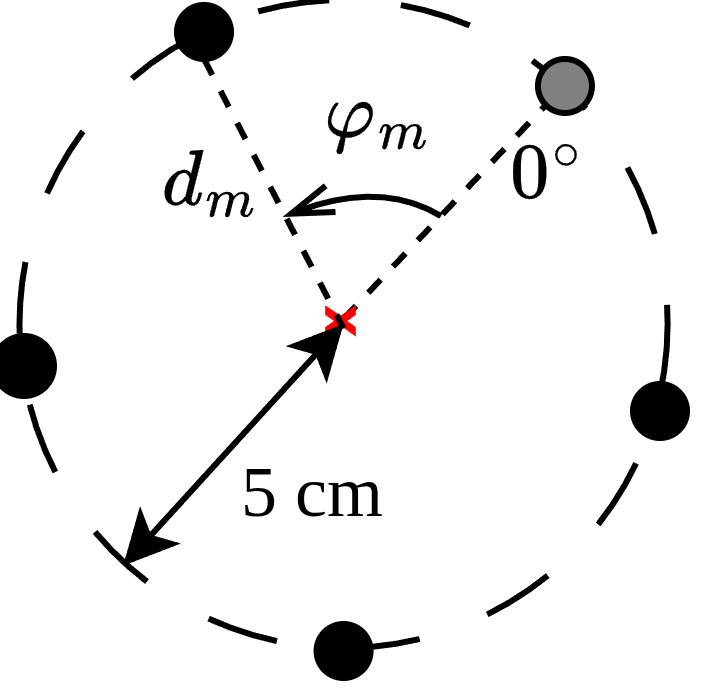}}
 \vspace{0.15cm}
  \centerline{(b) Circular}
    \label{figb}
\end{minipage}
 \caption{Considered random and circular array geometries with $M=5$ microphones. Gray circle denotes the reference microphone. Red cross denotes the array centroid.}
\label{fig:micgeo}
\end{figure}

To evaluate the proposed systems, acoustic environments were simulated using the pyroomacoustics library \cite{Scheibler2018pyroomacoustics}, with room dimensions and reverberation times ($T_{60}$) uniformly sampled from the ranges in Fig.~\ref{fig:setups} (right). To assess target speaker extraction performance, circular and random arrays with $M=5$ microphones were considered (see Fig.~\ref{fig:micgeo}). These arrays were positioned randomly in the horizontal plane at a height of 1.6~m with a random rotation. For the circular arrays, the microphones were evenly spaced along a circle with a diameter of $10$~cm. For the random arrays, microphone positions were sampled uniformly within a 10 cm × 10 cm region with a minimum microphone spacing of 2 mm, and a reference microphone was randomly chosen. As illustrated in Fig.~\ref{fig:setups} (left), two to six speakers were placed at a height of 1.6~m, ensuring a minimum $20^\circ$ angular separation between the target and all interfering speakers. The remaining angular space outside this target exclusion zone was divided into equal sectors corresponding to the number of interferers, with the DOA of each interferer randomly sampled within a distinct sector. Additionally, all speakers were positioned at least 50~cm apart from one another. Acoustic datasets were generated using 16~kHz speech signals from the Wall Street Journal (WSJ0) corpus \cite{wsj0data}, which was divided into disjoint speaker sets following the split defined in \cite{tesch2024ssf}. The training and validation sets comprise 30\,h and 1\,h of data, respectively. Similarly, 1\,h of data was generated for each array geometry for evaluation. The mixture signals were generated such that the reverberant signal-to-interference ratio at the reference microphone was between -5\,dB and 10\,dB. In addition, diffuse noise synthesized from the QUT noise dataset \cite{dean10_interspeech} using the ANF generator \cite{anf} was added such that the signal-to-noise ratio at the reference microphone was 15\,dB.
\subsection{Network and Training}
Following the framework in \cite{li2026flexiblemultichanneltargetspeaker}, the STFT is computed using a frame length of 512 (corresponding to 32~ms) with 50\% overlap and a square-root Hann window for both analysis and synthesis. The network architecture for all considered systems is the same as that of the GCSSF-MPE system in \cite{li2026flexiblemultichanneltargetspeaker}, except for the number of encoder output channels, which matches the dimensions of the input geometry-conditioning features. Specifically, for the proposed $\mathbf{P}_{\text{TSV}}$ and $\mathbf{P}_{\text{Mask}}$ features, the output channels of the three 1D convolutional (Conv1d) layers are set to 64, 512 and 1024, respectively. For the baseline GCSSF-MPE system, the positional encoding hyperparameters are set to $\sigma = 4$ and $Q = 514$, and the Conv1d output channels are adjusted to 64, 1024, and 514. This adjustment results in a comparable number of parameters in  GCSSF systems for a fair evaluation.

The training procedure follows the strategy in \cite{li2026flexiblemultichanneltargetspeaker}. All networks were trained to minimize a joint time- and frequency-domain loss function\cite{tesch2024ssf}, i.e.,
\begin{equation}
\mathcal{L}\hspace{-0.5mm}=\hspace{-0.5mm}\frac{\beta}{N} \hspace{-1mm}\sum_{n=1}^{N}\hspace{-0.5mm}|x_1(n)\hspace{-0.2mm} -\hspace{-0.2mm} \hat{x}_1(n)| \hspace{-0.1mm} + \hspace{-0.1mm}\frac{1}{KL}\hspace{-1mm}\sum_{l=1}^{L}\hspace{-0.7mm}\sum_{k=1}^{K}\hspace{-0.5mm}\left| |X_1(k,l)| \hspace{-0.8mm}- \hspace{-0.8mm}|\hat{X}_1(k,l)| \right|, \hspace{-1mm}\label{loss_func}\end{equation}
where $x_1(n)$ denotes the reverberant time-domain target speech signal at the reference microphone, and $\hat{x}_1(n)$ denotes its estimate reconstructed via inverse STFT and overlap-add. Here, $n$ is the time sample index, $N$ is the total number of samples, and the scalar weighting factor is set to $\beta = 10$. All networks were trained on random arrays, while performance was evaluated on circular and random arrays. To improve robustness against target DOA errors and microphone position mismatches, random perturbations were introduced during training: the target DOA was corrupted by an angular error drawn from $\mathcal{U}(-4^\circ, 4^\circ)$, and each microphone position was perturbed by an error drawn from $\mathcal{U}(-4~\text{mm}, 4~\text{mm})$. All networks were trained for 500 epochs with a batch size of 16 using the Adam optimizer (initial learning rate 0.001). A multi-step scheduler decayed the learning rate by a factor of 0.75 every 50 epochs up to epoch 400, and gradients were clipped to a maximum $l_2$-norm of 1 to ensure stable convergence.
\begin{table*}[t!]	
\vspace{-2ex}
\setlength{\tabcolsep}{8pt}
\centering
\caption{{Average SI-SDR (dB), PESQ, ESTOI, and DNSMOS overall scores (with standard deviations) for the unprocessed microphone signals, the baseline systems (SSF and GCSSF-MPE) and the proposed GCSSF systems using explicit spatial features (GCSSF-TSV and GCSSF-Mask). All systems are trained on random arrays and evaluated on both random and circular arrays.}}
\label{tab: comparebaseline_merged}
\small
\vspace{1mm}
\begin{tabular}{c|cccc|cccc}
    \toprule
    \multirow{2}{*}{System} & \multicolumn{4}{c|}{Random} & \multicolumn{4}{c}{Circular} \\
    & SI-SDR & PESQ & ESTOI & DNSMOS & SI-SDR & PESQ & ESTOI & DNSMOS \\
    \hline
    Unprocessed & \makecell{2.21 \\ ($\pm\,$4.24)} & \makecell{1.27 \\ ($\pm\,$0.19)} & \makecell{0.52 \\ ($\pm\,$0.14)} & \makecell{1.18 \\ ($\pm\,$0.19)} & \makecell{2.19 \\ ($\pm\,$4.28)} & \makecell{1.27 \\ ($\pm\,$0.20)} & \makecell{0.52 \\ ($\pm\,$0.14)} & \makecell{1.21 \\ ($\pm\,$0.22)} \\
    \hline
    SSF \cite{tesch2024ssf} & \makecell{6.78 \\ ($\pm\,$5.26)}  & \makecell{1.79 \\ ($\pm\,$0.49)} & \makecell{0.66 \\ ($\pm\,$0.16)} & \makecell{1.85 \\ ($\pm\,$0.56)} & \makecell{7.00 \\ ($\pm\,$5.25)}  & \makecell{1.80 \\ ($\pm\,$0.50)} & \makecell{0.66 \\ ($\pm\,$0.16)} & \makecell{1.86 \\ ($\pm\,$0.56)} \\
    GCSSF-MPE \cite{li2026flexiblemultichanneltargetspeaker} & \makecell{7.77 \\ ($\pm\,$5.01)} &\makecell{1.96 \\ ($\pm\,$0.55)} & \makecell{0.69 \\ ($\pm\,$0.16)} & \makecell{1.95 \\ ($\pm\,$0.58)} & \makecell{7.79 \\ ($\pm\,$4.93)} &\makecell{1.95 \\ ($\pm\,$0.55)} & \makecell{0.69 \\ ($\pm\,$0.15)} & \makecell{1.93 \\ ($\pm\,$0.57)} \\
    \hline
    GCSSF-TSV (\textbf{Proposed}) & \makecell{\textbf{9.43} \\ ($\pm\,$3.96)} & \makecell{\textbf{2.27} \\ ($\pm\,$0.55)} & \makecell{\textbf{0.75} \\ ($\pm\,$0.12)} & \makecell{\textbf{2.07} \\ ($\pm\,$0.59)} & \makecell{\textbf{9.87} \\ ($\pm\,$3.52)} & \makecell{\textbf{2.33} \\ ($\pm\,$0.54)} & \makecell{\textbf{0.77} \\ ($\pm\,$0.11)} & \makecell{\textbf{2.08} \\ ($\pm\,$0.59)} \\
    GCSSF-Mask (\textbf{Proposed}) & \makecell{8.45 \\ ($\pm\,$4.33)} & \makecell{2.10 \\ ($\pm\,$0.54)} & \makecell{0.72 \\ ($\pm\,$ 0.14)} & \makecell{1.99 \\ ($\pm\,$0.60)} & \makecell{8.86 \\ ($\pm\,$4.00)} & \makecell{2.17 \\ ($\pm\,$0.54)} & \makecell{0.74 \\ ($\pm\,$ 0.13)} & \makecell{2.02 \\ ($\pm\,$0.58)} \\
    \bottomrule                                     
\end{tabular}	
\end{table*}
\subsection{Experimental Results}
In this section, we present an experimental evaluation of the proposed GCSSF system utilizing explicit spatial features. First, we compare their target speaker extraction performance against the baseline SSF and GCSSF-MPE systems. Second, we analyze their robustness against inaccuracies in the input target DOA.
Target extraction performance is evaluated using three standard intrusive metrics: the scale-invariant signal-to-distortion ratio (SI-SDR) \cite{sisdr}, the wide-band perceptual evaluation of speech quality (PESQ) \cite{Rix2001PESQ}, and the extended short-time objective intelligibility (ESTOI) \cite{5495701}. For these metrics, the reverberant target speech component at the reference microphone is used as the ground-truth reference signal. Furthermore, to evaluate perceptual speech quality without a reference, the overall deep noise suppression mean opinion score (DNSMOS) \cite{9414878} is reported.

Table \ref{tab: comparebaseline_merged} compares the performance of the proposed systems (GCSSF-TSV and GCSSF-Mask) against the baseline systems (SSF and GCSSF-MPE) for random and circular arrays. Compared to the baseline SSF system, the GCSSF-MPE system achieves better performance across all metrics, indicating that mapping the target DOA and the microphone coordinates into sinusoidal embeddings helps the network to infer the relationship between inter-microphone phase differences, target DOA, and array geometry. Despite these improvements, providing the network with explicit spatial features yields significantly better results. Both proposed systems, GCSSF-TSV and GCSSF-Mask, consistently outperform the baseline systems across all metrics. 
Because the explicit spatial features are derived from acoustic propagation principles, they provide a highly informative representation that explicitly connects the inter-microphone phase differences in the multichannel signals with the target DOA and the array geometry. This simplifies the network's learning task, yielding a fundamental performance advantage.
Most notably, the GCSSF-TSV system achieves the best overall performance. On random arrays, it delivers substantial improvements of 1.66\,dB in SI-SDR, 0.31 in PESQ, 0.06 in ESTOI, and 0.12 in DNSMOS over the GCSSF-MPE baseline.
Although the spatial mask $\mathbf{P}_{\text{Mask}}$ incorporates additional input multichannel signal and indicates the presence of the target speaker, it performs worse than the target steering vector $\mathbf{P}_{\text{TSV}}$. This is likely because the TDOA compensation assumes direct-path propagation; consequently, reverberation degrades the phase alignment, leading to an inaccurate mask that suppresses time-frequency bins containing valid target speech. Furthermore, $\mathbf{P}_{\text{Mask}}$ is aggregated into a single-channel representation. Unlike the multichannel feature $\mathbf{P}_{\text{TSV}}$, it does not preserve inter-microphone phase information across the array, limiting the spatial information available to the network.
Evaluation on circular arrays demonstrates similar results to random arrays.

\begin{figure}[t]

\centering
\includegraphics[width=0.98\linewidth]{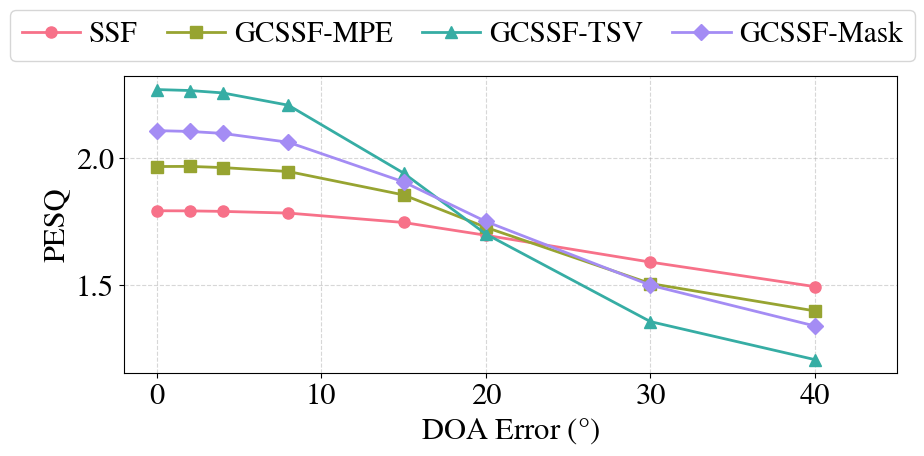}

\caption{Average PESQ of the baseline systems (SSF and GCSSF-MPE) and the proposed GCSSF systems using explicit spatial features (GCSSF-TSV and GCSSF-Mask) for different target DOA errors. All systems are trained and evaluated on the random arrays.} 
\label{fig:DOA_curve}
\end{figure}

Aiming at investigating the robustness against target DOA errors, Fig.~\ref{fig:DOA_curve} shows the average PESQ scores of all considered systems evaluated on random arrays for different target DOA errors. The results demonstrate that the proposed systems (GCSSF-TSV and GCSSF-Mask) maintain a clear performance advantage over the baseline systems for target DOA errors up to $20^\circ$. As the target DOA error increases beyond $20^\circ$, the PESQ scores of the proposed systems are lower than those of the baseline systems.  Because the proposed features are explicitly derived from the input target DOA, significant target DOA errors introduce severe phase mismatches and incorrect signal alignments in the target steering vector and the spatial mask. Consequently, the network is explicitly guided to focus on an incorrect DOA, suppressing the true target speaker.
In contrast, the baseline GCSSF-MPE system does not rely on acoustic propagation principles, thereby avoiding the severe performance degradation caused by signal misalignments under large target DOA errors. Overall, while the proposed spatial features introduce a sensitivity to large target DOA errors, they achieve better target speaker extraction performance for target DOA errors up to $20^\circ$.

\section{Conclusions}
\label{sec:conclusions}
In this paper, we addressed the limitations of abstract positional encodings in geometry-conditioned target speaker extraction by utilizing spatial features derived from acoustic propagation principles: the target steering vector (TSV) and the GCC-PHAT spatial mask. Unlike the baseline GCSSF-MPE system, which implicitly infers physical relationships from abstract encodings, the proposed systems use these explicit features to directly connect multichannel phase differences, the target DOA, and the array geometry. Experimental results demonstrate that the proposed systems consistently outperform the SSF and GCSSF-MPE baselines on both random and circular arrays. Furthermore, a sensitivity analysis on target DOA errors shows that the proposed systems maintain better target speaker extraction performance for target DOA errors up to $20^\circ$.

\bibliographystyle{IEEEbib}
\bibliography{strings,refs}

\end{document}